\documentclass[twocolumn,showpacs,preprintnumbers,amsmath,amssymb,prl]{revtex4}

\usepackage{graphicx}
\usepackage{dcolumn}
\usepackage{bm}
\usepackage{color}

\begin{document}

\preprint{}

\title{Intrinsic defects in silicon carbide LED as a perspective room temperature single photon source in near infrared}

\author{F.~Fuchs$^{1}$}
\author{V.~A.~Soltamov$^{2}$}
\author{S.~V\"{a}th$^{1}$}
\author{P.~G.~Baranov$^{2}$}
\author{E.~N.~Mokhov$^{2}$}
\author{G.~V.~Astakhov$^{1}$} 
\email[E-mail:~]{astakhov@physik.uni-wuerzburg.de}
\author{V.~Dyakonov$^{1,3}$}

\affiliation{$^1$Experimental Physics VI, Julius-Maximilian University of W\"{u}rzburg, 97074 W\"{u}rzburg, Germany \\
$^2$Ioffe Physical-Technical Institute, 194021 St.~Petersburg, Russia\\
$^3$Bavarian Center for Applied Energy Research (ZAE Bayern), 97074 W\"{u}rzburg, Germany}

\begin{abstract}
Generation of single photons has been demonstrated in several systems. However, none of them satisfies all the conditions, e.g. room temperature functionality, telecom wavelength operation, high efficiency, as required for practical applications.  Here, we report the fabrication of light emitting diodes (LEDs) based on intrinsic defects in silicon carbide (SiC). To fabricate our devices we used a standard semiconductor manufacturing technology in combination with high-energy electron irradiation. The room temperature electroluminescence (EL) of our LEDs reveals two strong emission bands in visible and near infrared (NIR), associated with two different intrinsic defects. As these defects can potentially be generated at a low or even single defect level, our approach can be used to realize electrically driven single photon source for quantum telecommunication and information processing.
\end{abstract}

\date{\today}


\maketitle

Robust and cheap light sources emitting single photons on demand are at the heart of many demanding optical technologies \cite{Review1,Review2}. Single photon emission has been demonstrated in a variety of systems, including atoms \cite{SP-Atom}, ions \cite{SP-Ion}, molecules \cite{SP-Molecule,SP-Molecule2,SP-Molecule3}, quantum dots (QDs) \cite{SP-QD,LED-QD} and color centers in diamond \cite{SP-NV,SP-SiV}. The most significant progress has been achieved for QDs \cite{QD-Cavity,QD-Ent,QD-BB84}, however, the necessity to use cryogenic temperatures and high inhomogeneity (the emission wavelength is individual for each QD) make this system impractical. Electrically driven single photon sources in visible have also been demonstrated using nitrogen-vacancy (NV) centers in diamond  \cite{LED0-NV,LED-NV}, but the compatibility of this system with the present-day integrated circuits manufacturing is not obvious. 

\begin{figure}[btp]
\includegraphics[width=.45\textwidth]{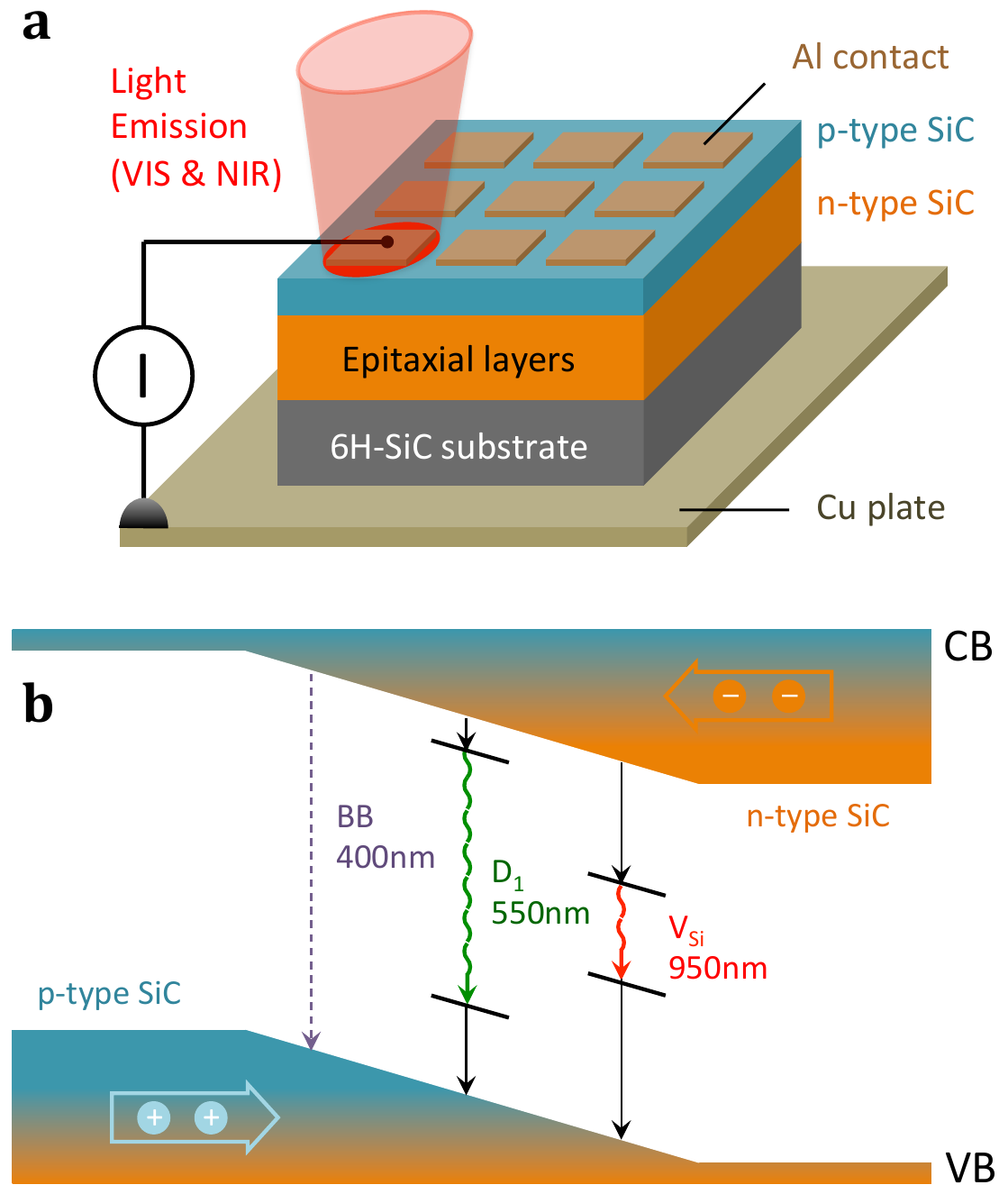}
\caption{(a) A scheme of the SiC LED. (b) Electron-hole recombination through the $\mathrm{D_1}$ and $\mathrm{V_{Si}}$ defects results in the $550 \, \mathrm{nm}$ and $950 \, \mathrm{nm}$ emission bands, respectively. The radiative band-to-band recombination (BB) at $400 \, \mathrm{nm}$ is inefficient because SiC is an indirect bandgap semiconductor.} \label{fig1}
\end{figure}

The operation principle of single photon sources is based on the quantum mechanical properties of a single two-level system. When a single photon is desired, this system is put into the excited state by an external stimulus, and a single photon is emitted upon relaxation into the ground state. A perspective approach to fabricate an efficient, room temperature single photon source based on this principle is to use color centers in semiconductors.  In our work, we exploit two  defect centers in SiC, the so-called $\mathrm{D_1}$ defect \cite{D1-Obser} and the silicon vacancy ($\mathrm{V_{Si}}$) defect \cite{SiC-SiliconVacancy}, making two-color LED [Fig.~\ref{fig1}(a)]. 

Remarkably, the $\mathrm{V_{Si}}$ defects in SiC comprise the technological advantages of semiconductor quantum dots and the unique quantum properties of the NV defects in diamond \cite{SiC-ResonantControl}. In particular, $\mathrm{V_{Si}}$ spin qubits can be optically initialized and read out, and, therefore, our demonstration of room temperature EL from $\mathrm{V_{Si}}$ defects is an important step towards realization of all-electrical control of $\mathrm{V_{Si}}$ spins. Further, the $\mathrm{V_{Si}}$ EL reveals a broad-band emission spectrum in NIR ($850-1050 \, \mathrm{nm}$), where the absorption of  silica glass optical fibers is relatively weak.  While  this spectrum range is still below the telecom window ($1.3 \, \mathrm{\mu m}$), it can be changed in the direction of long wavelengths by proper choosing over family of  $\mathrm{V_{Si}}$-related defects in different SiC polytypes. Alternatively, the frequency conversion of NIR photons to a telecom wavelength can be applied \cite{DownConversion1,DownConversion2}. Therefore, the integration of defect-based SiC LEDs with existing telecommunication infrastructure seems feasible.  

\begin{figure}[btp]
\includegraphics[width=.49\textwidth]{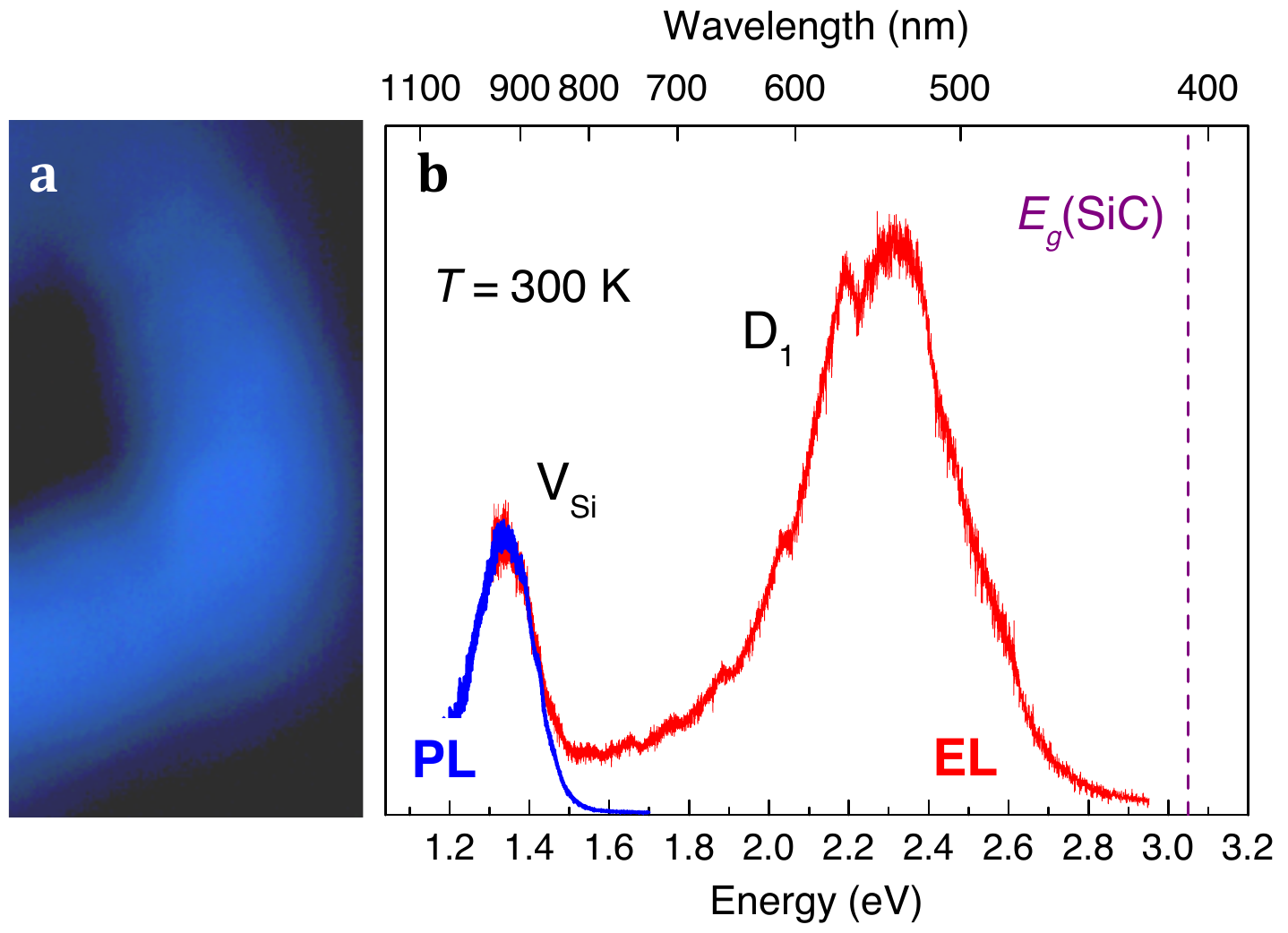}
\caption{(a) An image of the luminous LED around an Al contact. (b) Electroluminescence (EL) spectrum of the SiC LED and photoluminescence (PL) spectrum of the reference SiC sample recorded at room temperature. The PL spectrum is excited by a He-Ne laser with $E_{exc} = 1. 96 \, \mathrm{eV}$ ($633 \, \mathrm{nm}$). The bandgap of 6H-SiC is $E_g (\mathrm{SiC})= 3.05  \, \mathrm{eV}$.} 
\label{fig2}
\end{figure}

SiC with highly developed device technologies (e.g. MOSFETS, MEMS, sensors) is a very attractive material for practical applications. SiC is also known as the material on which the first LED has been created \cite{First-LED}. Until the 90's, SiC was used for commercial yellow and blue LEDs, but later it was replaced with GaN. The reason is that SiC is an indirect bandgap semiconductor and as a consequence direct band-to-band radiative recombination is inefficient. For the same reason, recombination through defects is preferential in SiC [Fig.~\ref{fig1}(b)]. If one would isolate a single defect with radiative recombination channel, this property of SiC could be used to efficiently generate single photons. 

The LED structures used in our experiments were grown on a n-type 6H-polytype SiC substrate [Fig.~\ref{fig1}(a)]. First, an epitaxial 15-$\mathrm{\mu m}$-thick SiC layer was grown by the sublimation method. The layer is n-type with donor concentrations of $3 \times 10 ^{18} \, \mathrm{cm^{-3}}$ (N). It also contains Ga impurities ($2 \times 10 ^{18} \, \mathrm{cm^{-3}}$), serving as PL activation centers. The layer is followed by a p-type SiC layer of thickness  $5 \, \mathrm{\mu m}$ grown at a temperature of $2300 \, \mathrm{C^{\circ}}$ in Ar atmosphere in the presence of Al vapors (pressure $100 \, \mathrm{Pa}$).  This results in the concentration of Al acceptors of ca. $10 ^{20} \, \mathrm{cm^{-3}}$. In order to generate intrinsic defects at the p-n junction the samples were irradiated with  $0.9 \, \mathrm{MeV}$ electrons to a dose of $10 ^{18} \, \mathrm{cm^{-2}}$. After irradiation, the samples were annealed for 10 minutes in Ar atmosphere at a temperature of $1700 \, \mathrm{C^{\circ}}$. At the final stage, $0.4 \times 0.4 \, \mathrm{mm^2}$ Al contacts were deposited on the top of the p-type SiC layer.  

We mount LED samples on a Cu plate serving as the back electrode. Upon applying voltage between an Al contact and the Cu plate the luminescence glow is seen by the naked eye [Fig.~\ref{fig2}(a)]. The room temperature EL spectrum of one of our LEDs is presented in Fig.~\ref{fig2}(b). It consists of two broad emission bands, labeled as $\mathrm{D_1}$ and $\mathrm{V_{Si}}$. The corresponding recombination processes at the p-n junction are schematically shown in Fig.~\ref{fig1}(b). 

We now discuss the EL bands of  Fig.~\ref{fig2}(b) in details. Remarkably, the corresponding emission energies are significantly smaller than the bandgap of 6H-SiC ($3.05 \, \mathrm{eV}$). We, therefore, ascribe them to the defects in SiC. The emission in the spectral range $450-650 \, \mathrm{nm}$  is characteristic for the $\mathrm{D_1}$ defect \cite{D1-Obser}. The nature of this defect is still not clear -- several models have been proposed, including bound-exciton-like center \cite{D1-1} and first-neighbor antisite pair $\mathrm{Si_{C} - C_{Si}}$ \cite{D1-2}. The second emission band in the NIR spectral range $850-1050 \, \mathrm{nm}$ coincides with the photoluminescence (PL) spectrum of the silicon vacancy defects $\mathrm{V_{Si}}$ \cite{Ilyin1981} in the reference 6H-SiC bulk sample. 

\begin{figure}[btp]
\includegraphics[width=.45\textwidth]{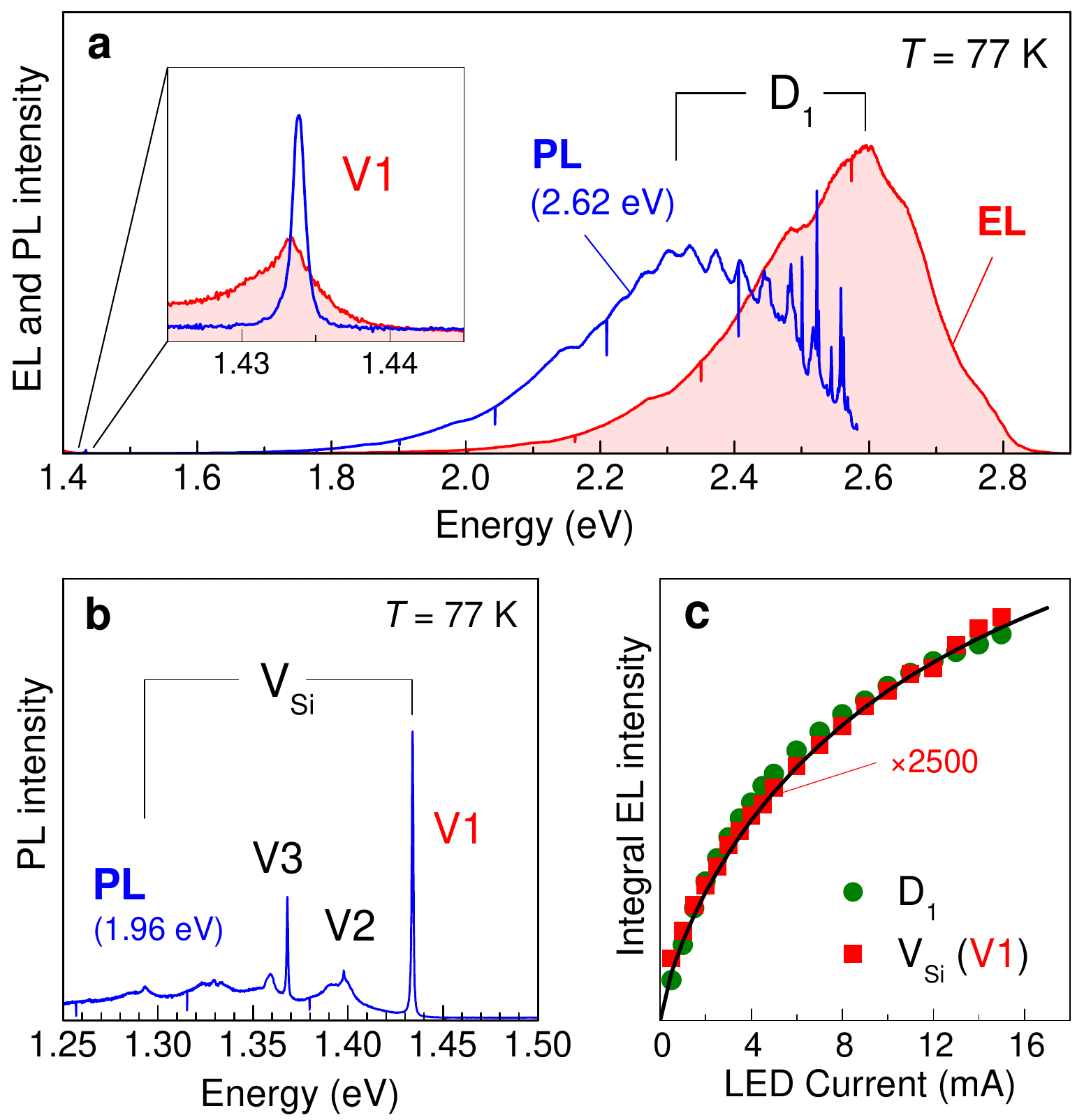}
\caption{EL and PL spectra of the SiC LED recorded at  $T = 77 \, \mathrm{K}$. (a) Comparison of the EL (shaded area) and PL spectra under excitation with an energy $E_{exc} = 2. 62 \, \mathrm{eV}$ ($473 \, \mathrm{nm}$). Inset: The same, but shown in the spectral range where the strongest $\mathrm{V_{Si}}$ ZPL (V1) is expected. (b) PL spectrum obtained under excitation with  a He-Ne laser with $E_{exc} = 1. 96 \, \mathrm{eV}$ ($633 \, \mathrm{nm}$).  The V1, V2 and V3 ZPLs characteristic for the $\mathrm{V_{Si}}$ defects in SiC are clearly seen.  (c) Integral intensity of the  $\mathrm{V_{Si}}$ and $\mathrm{D_1}$ mission bands [the shaded areas in (a)] as a function of LED current. The solid line is a fit (see text for details).} 
\label{fig3}
\end{figure}

To prove this interpretation we repeat the experiment of Fig.~\ref{fig2}(b) at a temperature of   $77 \, \mathrm{K}$ [see Fig.~\ref{fig3}(a)], when the spectroscopic features individual for each defect can be resolved. In this low-temperature experiment, the samples are kept in vacuum and the Cu plate is also used as a cold finger. The results are summarized in Fig.~\ref{fig3} and below we discuss them in details. 

First, we demonstrate the presence of $\mathrm{V_{Si}}$ defects in our LED structures. Figure~\ref{fig3}(b) shows photoluminescence (PL) spectrum recorded under excitation with the energy  $E_{exc} = 1. 96 \, \mathrm{eV}$, which is below the $\mathrm{D_1}$ emission energy. Three zero-phonon lines (ZPLs) at $1.368 \, \mathrm{eV}$, $1.398 \, \mathrm{eV}$ and $1.434 \, \mathrm{eV}$ are the well known fingerprint of the $\mathrm{V_{Si}}$ defects in 6H-SiC \cite{Wagber2000}. These three ZPLs originate from three nonequivalent crystallographic sites in this SiC polytype and are frequently labeled as V3, V2, and V1, respectively. The highest ZPL intensity is observed for $\mathrm{V_{Si}(V1)}$.

Second, we demonstrate that the $\mathrm{V_{Si}(V1)}$ defect can be electrically driven. Figure~\ref{fig3}(a) shows EL spectrum recorded at $T = 77 \, \mathrm{K}$. In contrast to room temperature [Fig.~\ref{fig2}(b)], the $\mathrm{D_1}$ emission dominates in the spectrum. The reason is the much higher concentration of $\mathrm{D_1}$ defects than of $\mathrm{V_{Si}}$ defects. However, the activation energy of the $\mathrm{D_1}$ defect is relatively small ($57 \, \mathrm{meV}$  \cite{D1-Donor}), leading to the intensity increase with lowering temperature. On the other hand, the activation energy of the $\mathrm{V_{Si}}$ defects is much higher and their intensity weakly depends on temperature. Indeed, we observe the characteristic $\mathrm{V_{Si}(V1)}$ ZPL at $1.434 \, \mathrm{eV}$ in the EL spectrum [the inset of Fig.~\ref{fig3}(a)]. 

Third, we verify that the electrical excitation of the $\mathrm{V_{Si}(V1)}$ defect shown in the inset of Fig.~\ref{fig3}(a) is not due to the re-emission process via $\mathrm{D_1}$. We excite PL with the energy of $2.62 \, \mathrm{eV}$, coinciding with the maximum of the $\mathrm{D_1}$ EL band [Fig.~\ref{fig3}(a)]. The laser intensity per area is several orders of magnitude higher than that of the $\mathrm{D_1}$ emission, but no significant enhancement of the $\mathrm{V_{Si}(V1)}$ PL is observed. This means that while the reemission may potentially take place, it is inefficient as compared to electrical excitation. Therefore, we conclude, the recombination of electrically injected electron and holes is responsible for the $\mathrm{V_{Si}}$ EL, as schematically shown in Fig.~\ref{fig1}(b). 

Finally, we present an input-output characteristic of one of our LED devices [Fig.~\ref{fig3}(c)]. A clear tendency to saturation of the emission intensity $P$ with injection current $I$ is seen. This behavior can be well described as $P \propto (1 + I_0 / I)^{- \alpha}$ [the solid line in Fig.~\ref{fig3}(c) corresponds to $\alpha = 0.75$ ]. The characteristic saturation current $I_0 = 17 \, \mathrm{mA}$ is higher than that in QD-based single photon LEDs \cite{QD-II} and comparable with that in NV-based single photon LEDs \cite{LED-NV}. 

In conclusion, we generated intrinsic defects in SiC devices and demonstrated that these defects can be electrically driven, resulting in the efficient EL with emission energies well below the SiC bandgap. Our LEDs are two-color in a sense that they show two spectrally different emission bands associated with different defects. The $\mathrm{D_1}$ defects show  EL in visible, which is intense at low temperatures but quenches with rising temperature. The  $\mathrm{V_{Si}}$ defects emit in NIR even at room temperature. By varying the irradiation dose one can control defect concentration, which  should allow to isolate single defects, similar to single NV centers in diamond or single semiconductor QDs. Because isolated defects are ideal single photon emitters, our findings open a new way to fabricate cheap and robust LEDs emitting single photons on demand in NIR.



\end{document}